\documentstyle[11pt,a4,psfig,cite]{article}
\let\oldsubsec=\subsection\let\oldbfs=\bfseries
\def\subsection#1{\def\bfseries{\it}\oldsubsec{#1}
\let\bfseries=\oldbfs}
\def\beq{\begin{equation}}
\def\eeq{\end{equation}}
\def\eq#1{Eq.~(\ref{#1})}
\def\k{{\bf k}}\def\q{{\bf q}}\def\r{{\bf r}}\def\Qh{\hat{\bf Q}}
\def\F{{\bf F}}\def\P{{\bf P}}\def\Q{{\bf Q}}\def\R{{\bf R}}
\def\a{\alpha}\def\s{\sigma}\def\t{\tau}
\def\w{\omega}\def\dw{\Delta\omega}
\def\sqw{S_i(\Q,\w)}\def\yqt{Y(\Q,t)}\def\jqy{J(\Qh,y)}
\def\ncs{neutron Compton scattering}
\def\ia{impulse approximation}
\def\dos{density of states}
\def\cp{Compton profile}
\def\etal{{\it et al.}}
\def\av#1{\langle{#1}\rangle}
\def\frac#1/#2{\leavevmode\kern.1em
  \raise.5ex\hbox{\the\scriptfont0 #1}\kern-.1em
  /\kern-.15em\lower.25ex\hbox{\the\scriptfont0 #2}}
\def\half{\frac1/2}

\begin{document}
\title{Neutron Compton Scattering}
\author{G.~I.~Watson\\
Rutherford Appleton Laboratory\\
Oxfordshire OX11 0QX, UK}
\date{28 May 1996}
\maketitle

\begin{abstract}
Neutron Compton scattering measurements have the potential to provide
direct information about atomic momentum distributions and adiabatic
energy surfaces in condensed matter.  First applied to measuring the
condensate fraction in superfluid helium, the technique has recently
been extended to study a variety of classical and quantum liquids and
solids.  This article reviews the theoretical background for the
interpretation of neutron Compton scattering, with emphasis on studies
of solids.
\end{abstract}

\section{Introduction}
Neutron Compton scattering is set apart from other branches of neutron
scattering by the magnitude of energy and momentum transfers involved,
typically in excess of 1 eV and 30 \AA${}^{-1}$, respectively.  Neutron
scattering is renowned as a sensitive probe of collective properties in
condensed matter, but in this extreme range of energies and momenta---%
the impulse limit---it is single-particle properties that are probed.  The
scattering occurs so rapidly, compared with the time scales of atomic
motion in the sample, that the measured response is rather simply related
to the equilibrium momentum distribution of the atoms.

Thus, the term \ncs\ refers, not to a distinct type of scattering, but
to the usual neutron scattering cross-section in a limiting range of
parameters.  In the impulse limit, the scattered intensity, as a
function of energy for fixed momentum transfer $\Q$, consists of a peak
centred at $E_R=\hbar^2Q^2/2M$, the energy of recoil of a
stationary nucleus on colliding with a neutron.  The dependence of
the peak position on the mass $M$ of the struck atom implies that the
scattering from different atoms appear at different energies.  This
ability to separate contributions from different atomic species (and
isotopes) is a useful feature of the technique.

Doppler broadening results in a recoil peak whose width is proportional
to the mean kinetic energy of the atoms, and whose detailed shape
depends on the distribution of atomic momenta.  The principal tool in
interpreting experiments is the impulse approximation, which predicts
a precise relationship between the scattering data, in the form of the
Compton profile, and the atomic momentum distribution.  This gives
an opportunity to compare results of experiments with predictions of
realistic theoretical models of microscopic properties.

As the name suggests, \ncs\ is closely related to experimental
techniques in other branches of physics \cite{silv89a}.  Compton scattering
\cite{will77,coop85,kane92}, i.e.~inelastic scattering of X-rays, gamma
rays or (10-60 keV) electrons provides information on electron momentum
distributions, an example being the direct observation of the Fermi--%
Dirac distribution for conduction electrons in metals.  In nuclear physics,
an established technique for probing the momentum distribution of nucleons
in nuclei is deep inelastic scattering \cite{anto88,sick89} of protons or
electrons with energies of 100--1000 MeV and de~Broglie wavelengths of the
order of nuclear radii.  At still higher energies (10 GeV and above), deep
inelastic scattering of electrons, muons or neutrinos \cite{tayl91,%
kend91,frie91} probes the internal structure of nucleons and has been
a key experiment in the confirmation of the existence of quarks.  By
analogy with these techniques, \ncs\ is sometimes called deep inelastic
neutron scattering.

Neutron Compton scattering was first suggested \cite{hohe66} as a
technique for measuring of the condensate fraction in superfluid
${}^4$He, and this has been an active field for three decades (see
\cite{silv89a,sosn91,glyd94a} and references therein).  Recently, the
development of spallation neutron sources, which have much higher
intensities than reactor sources in the incident energy range required
for \ncs, have opened up new applications.  Recent experiments include
studies of condensed noble gases \cite{peek91,frad94a,frad94b,azua95a},
metals \cite{evan94,fult94}, normal liquid ${}^4$He \cite{azua95a,%
ande94,andr94,bafi95}, liquid ${}^3$He \cite{azua95b}, solid ${}^4$He
\cite{bafi95}, superfluid ${}^3$He--${}^4$He mixtures \cite{wang94},
and molecular hydrogen and deuterium \cite{maye93,andr95,bafi96a,bafi96b}.

The extreme momentum transfers now experimentally reachable imply
that, in some cases, the corrections to the \ia\ are essentially
negligible.  This suggests the possibility of direct and model-independent
extraction of momentum distributions from Compton profiles, in contrast to
the usual approach of fitting theoretical predictions to the experimental
profiles.  Though its feasibility has not yet been demonstrated in
practice, this approach, if successful, could establish \ncs\ as the
only technique capable of measuring atomic momentum distributions directly.
A particularly interesting case is where the motion of the atom of
interest is well described by an effective (Born--Oppenheimer) potential,
or adiabatic energy surface.  An example is a proton bound in a heavy
lattice.  In this case, the momentum distribution is the squared amplitude
of the Fourier transform of the proton wave function, and from it, the
potential energy function can be extracted \cite{reit85a,reit89}.

The information obtainable in \ncs\ is to some extent complementary
to that from diffraction experiments.  The former measures the
Fourier transform of a {\it time averaged} density; the latter
the {\it instantaneous} momentum density.  Thus, for example, a \ncs\
experiment on protons in a double-well potential of a hydrogen bond
\cite{reit85a,reit89} could, in principle, distinguish between a wave
function with amplitude in both wells, and a statistical mixture of
states in which the proton is localized in one well or the other.

This article aims to summarize the theoretical framework for the
interpretation of \ncs\ experiments, with an emphasis on applications to
solids.  A thorough discussion of the physical principles underlying the
impulse approximation is given in Sec.~2, based on the central concept
of the scattering time.  Sec.~3 addresses the problem of extracting
information from the Compton profile. In the final section we discuss
briefly the potential of the direct inversion approach.  Although
some examples are drawn from published experimental studies, no attempt
is made here to review current experimental activity.  For further
background material, see Refs.~\cite{soko94,love95}.

\section{The \ia}
We begin with an expression for the quantity which is measured in
experiments, namely the partial differential cross section for the
scattering of a beam of neutrons by a collection of atomic nuclei.
Although the nucleon--nucleon interaction is very strong, it is
sufficiently short-ranged that the scattering is a weak perturbation
of the incident wave.  Therefore, with the aid of a Fermi pseudo-potential,
the scattering may be described in the first Born approximation, and as a
consequence the cross-section depends only on the changes in energy
and wave vector of the neutron.  We shall denote these by
\beq
\hbar\w={\hbar^2\over2m}(k^2-k'^2)\quad\hbox{and}\quad\Q=\k-\k',
\eeq
respectively, where $\k$ and $\k'$ are the incident and scattered
neutron wave vectors.  A second simplifying assumption appropriate
in the case of \ncs\ is that the spatial scale of the scattering
event, set by $1/Q$, is too small to detect correlations between
the positions of different nuclei, and hence that the scattering
may be described to a good approximation as incoherent.  This
approximation is particularly good for scattering from protons,
where the incoherent cross-section is larger than the coherent
cross-section by almost two orders of magnitude.  Under these
circumstances the contribution to the scattering cross-section from
nuclei of a particular species is proportional to the response
function \cite{love84}
\beq
\sqw={1\over2\pi\hbar}\int_{-\infty}^\infty dt\,e^{-i\w t}
\sum_j Y_j(\Q,t),\label{siqw}
\eeq
where
\beq Y_j(\Q,t)=\av{e^{-i\Q\cdot\R_j}e^{i\Q\cdot\R_j(t)}} \label{yj}\eeq
is the density-density correlation function corresponding to the nucleus
of atom $j$, whose position is represented by the quantum mechanical
operator $\R_j$.  The corresponding $\R_j(t)$ has a time dependence
in the Heisenberg representation,
\beq \R_j(t)=e^{iHt/\hbar}\R_je^{-iHt/\hbar}. \eeq
The angular brackets in \eq{yj} denote a thermal average of the
enclosed expression, as well as an implicit average over degrees of
freedom which are passive in the scattering process, such as nuclear
spin and neutron polarization states.

The physical content of the response function can be made more
transparent by passing to a somewhat less general expression.  The
thermal average in \eq{yj} involves, in principle, a sum over
expectation values with respect to the complete set of stationary
quantum states of the many-body Hamiltonian describing the nuclei
and other particles, and their interactions.  Let us suppose that
the motion of a particular nucleus can be represented by an effective
single-particle Hamiltonian which describes its interaction with
its environment, and a corresponding set of single-particle states.
We have in mind the case of a nucleus bound in a molecule, where
an effective potential is constructed using the Born--Oppenheimer
scheme.  Let us denote the effective single-particle states
by $|n\rangle$ and their energies by $E_n$.  The contribution to
the response function from a single nucleus reduces to
\beq
\sqw=\sum_{nn'}Z^{-1}e^{-E_n/k_BT}|\av{n|e^{-i\Q\cdot\R}|n'}|^2
\delta(\hbar\w-(E_{n'}-E_n)),\label{boundia}
\eeq
where $Z$ is a normalization factor (the thermodynamic partition function).
In this expression we may recognize a sum over transitions from
initial states $n$, weighted by a thermal Boltzmann factor, to final
states $n'$, of a transition probability.  The latter is given by
Fermi's golden rule as the product of a squared matrix element and
an energy-conserving delta function.  The scattering is represented
by the operator $e^{-i\Q\cdot\R}$ which couples the plane wave of
the neutron with the position of the nucleus.

Returning now to the general expression, \eq{siqw}, the form of the
response function as a sum of separate contributions from each
nucleus reflects the nature of the incoherent approximation, which
neglects correlations between different nuclei.  The \ia, which is
the main topic of this section, consists in a similar neglect of
{\it time} correlations in the motion.  To be specific, if the time
scale of the scattering event is much shorter than the characteristic
time of atomic motions in the sample, then the nuclei may be regarded
as free particles in so far as the scattering probability is concerned.
As a result, the latter depends only on the momentum of the nucleus in
its initial state.
Crudely speaking, then, we may summarize the approximations appropriate
for \ncs\ as follows: the Born approximation, that each neutron
scatters only once; the incoherent approximation, that each scattering
process involves only one nucleus; and the \ia, that the neutron remains
in the vicinity of the nucleus for a time too short to sense anything
except how fast the nucleus is moving.  This last approximation is the
most subtle of the three, and it will be discussed in detail now.

\subsection{The short time expansion}
Motivated by the idea that only short times are relevant, let us
suppose that the correlation function $\yqt$ for a single nucleus is
dominated by its behaviour for small $t$.  The time dependent Heisenberg
operator $\R(t)$ has a Taylor expansion,
\beq \R(t)=\R+(\P/M)t+\half(\F/M)t^2+\ldots, \label{shortim}\eeq
where $\P=M\dot\R=iM[H,\R]/\hbar$ is the momentum of the struck
nucleus which has mass $M$, and $\F$ is the force defined similarly.
All the operators in the expansion are evaluated at $t=0$.  We remark
that the identification of $M\dot\R$ with momentum is valid only in
the absence of velocity-dependent forces, i.e.~it is assumed that there
are no magnetic fields present, and that the motion of the nucleus is
non-relativistic.

Reserving for later the question of the precise criteria for validity
of the present mathematical procedure, let us examine the form of the
response function resulting from the neglect of terms of order $t^2$
and higher.  This amounts to an assumption that the struck particle is
effectively free so that $\F=0$.  With the aid of the operator identity
\beq e^{A+B}=e^Ae^Be^{-\half[A,B]}, \eeq
which holds when $[A,B]$ commutes with both $A$ and $B$, the correlation
function is found to reduce to
\beq
\yqt = \av{e^{i(\Q\cdot\P/M+\w_R)t}},\label{impy}
\eeq
where $\w_R=\hbar Q^2/2M$ is the (free atom) recoil frequency.
The exponent has an interpretation in terms of the kinematics of a
neutron--nucleus collision.  If the momentum of the nucleus is $\P$
before the collision, then it is $\P+\hbar\Q$ after the collision,
and energy conservation requires
\beq \w=[(\P+\hbar\Q)^2-P^2]/2M\hbar=\Q\cdot\P/M+\w_R.\label{cons} \eeq
In particular, $\hbar\w_R$ is the energy imparted to a stationary nucleus
by a collision with a neutron.

If the nucleus were indeed at rest before the scattering, \eq{impy}
would give  $\yqt=e^{i\w_Rt}$, and the response function would consist
of a delta function line at the recoil frequency.  In fact, the nucleus
will be in a quantum state having a distribution of initial momenta, and
the line will therefore be Doppler broadened.  Each possible initial
momentum results in a contribution to the scattering intensity at a
frequency given by the conservation constraint, \eq{cons}.  Specifically,
defining the momentum distribution by
\beq n(\q)=\av{\delta(\q-\P/\hbar)},\label{momdist} \eeq
the response function in the \ia\ is
\beq
\sqw=\hbar^{-1}\int n(\q)\delta(\w-\hbar\Q\cdot\q/M-\w_R)\,d\q.
\label{impsqw}
\eeq
This result, that the scattering cross-section is directly related
to the momentum distribution of the struck nuclei, is of central
importance in \ncs\ experiments.

It is important to emphasize that, although it represents a
single-particle response, the momentum distribution $n(\q)$ is
actually a property of the many-body system of all the nuclei and
their interactions.  In other words, the momentum distribution of
a single nucleus depends on its environment and therefore on the
behaviour of the system as a whole.  Let us examine again the special
case where the motion of a nucleus is given by a set of effective
single-particle states $|n\rangle$.  Here, \eq{momdist} becomes
\beq
n(\q)=(1/2\pi)^3\sum_nZ^{-1}e^{-E_n/k_BT}\left|\int e^{-i\q\cdot\r}
\av{\r|n}\,d\r\right|^2.\label{momft}
\eeq
At zero temperature, this reduces to the square modulus of the
Fourier transform of the ground state wave function.  Thus, there
is a direct relationship between the measured quantity, $\sqw$,
and the quantum mechanical wave functions of the nuclei in the
sample.

\subsection{The scattering time}
Looking at the short time expansion of the nucleus' position
operator, \eq{shortim}, we see that the first term neglected in the
derivation of the \ia\ is proportional to the operator $\F$ representing
the force experienced by the nucleus.  Denoting root mean square values
by an overbar, the corrections are expected to be small if
\beq
\overline{\F\cdot\Q}\,\t_s \ll\overline{\P\cdot\Q},\label{impcond}
\eeq
where $\t_s$ is a quantity with the dimensions of time, identified
as the time scale of the scattering process, or {\it scattering
time}.  At first sight, it might appear reasonable to relate $\t_s$
to the time taken for a neutron wave packet to pass the vicinity of
the nucleus.  The corollary, that the degree of coherence of the
neutron beam plays a role in deciding the validity of the \ia, is
largely erroneous, as will be demonstrated presently.

To establish the significance of the scattering time, let us adopt
the operational point of view that, since the aim is to approximate
the correlation function $\yqt$, the relevant time scale should
be obtained from $\yqt$ itself.  The response function $\sqw$, as we
have seen, consists in general of a peak centred at the recoil
frequency $\w_R$, with a certain width $\dw$.  It follows that
the general form of $\yqt$ is of an oscillatory function $e^{i\w_Rt}$,
modulated by a decreasing envelope of width $1/\dw$ (Fig.~1).
The fact that $\yqt$ goes to zero confirms our expectation from
physical arguments that knowledge of the struck atom's motion for a limited
time span is sufficient to predict the scattering response.  Accordingly,
we identify the scattering time $\t_s$ as the time scale for the
decay of the correlation function $\yqt$ to zero, which equals $1/\dw$,
the reciprocal of the recoil peak width \cite{glyd94a}.
If it happens that $\sqw$ is highly structured, i.e.~has features on
several frequency scales, this means that $\yqt$ depends on more than
one characteristic time.  The overall decay envelope is determined
by the longest of these characteristic times, and it follows that
$\t_s$ is the reciprocal of the width of the {\it narrowest} feature
in $\sqw$.

\def\captone{\small
Schematic form of the correlation function $\yqt$,
whose Fourier transform is proportional to the neutron scattering
cross section.  The frequency of the oscillations is the reciprocal
of the recoil frequency of the scatterer, and the amplitude falls
off on a scale of the scattering time $\t_s$.}
\begin{figure}
\centerline{\psfig{file=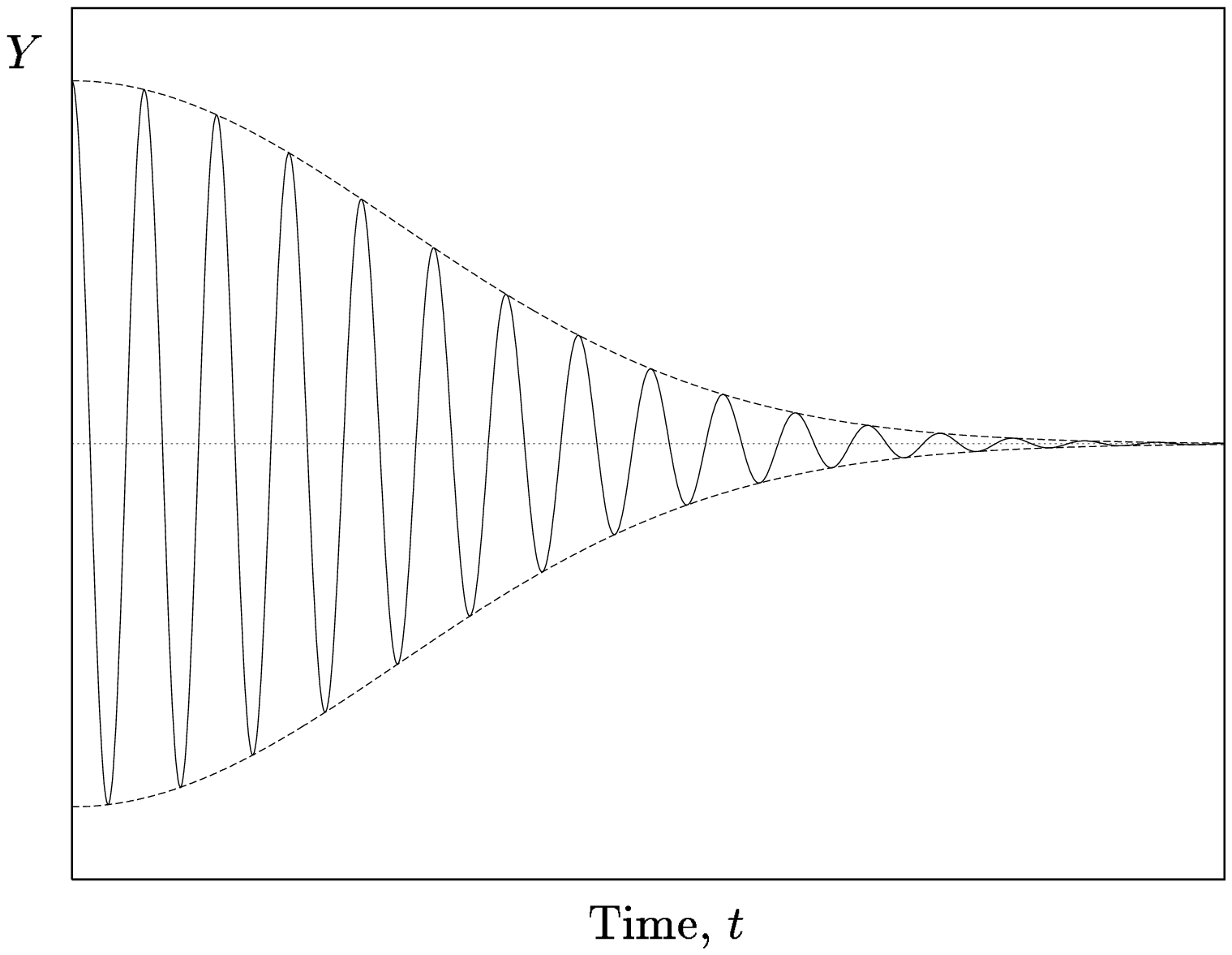,width=4truein}}
\caption{\captone}
\end{figure}

Strictly speaking, for scattering from solids the correlation function
$\yqt$ does not go to zero, but to $|\av{e^{i\Q\cdot\R}}|^2$,
corresponding to the amplitude of the elastic line in $\sqw$ \cite{love84}.
However, this constant component does not affect the argument: since
we are interested in the scattering at energies near $\w=\w_R$, we may
discard the elastic scattering and take $\yqt$ to consist only of
the inelastic part, which tends to zero as described.

With $\t_s$ defined in terms of the structure of the response function,
a rigorous determination requires, in principle, a detailed calculation
of $\sqw$ for the system of interest.  In the absence of a detailed
theory, some rough estimates can be made.  A convenient procedure
is to estimate $\t_s$ self-consistently using the \ia\ itself.  We
take for $\sqw$ a peak centred on $\w_R$ of width $\dw$.  According
to \eq{impsqw}, $\dw$ is proportional to the width of the distribution
of $P_Q$, the projection of momentum $\P$ along the direction of $\Q$,
and thus
\beq \t_s\sim{M\over Q\av{P_Q^2}^{1/2}}.\label{scatime} \eeq
The inverse dependence of the scattering time on $Q$ is an essential
feature.  It implies that, as anticipated on physical grounds, the
\ia\ becomes exact in the limit of large wave vector transfer.  This
statement is true provided the forces on the nuclei are always finite,
for if $\F$ can grow arbitrarily large the condition in \eq{impcond} is
not satisfied no matter how short the scattering time.  We shall
therefore not consider pathological cases, such as scattering from a
``hard-core'' fluid \cite{reit85b} or from a particle undergoing
Brownian motion \cite{datt85}, where corrections to the \ia\ do not
become negligible as $Q\to\infty$.

With our estimate, \eq{scatime}, for the scattering time, the
criterion for validity of the \ia\ reads
\beq \overline{F_Q}(1/Q) \ll\av{P_Q^2}/M, \label{selfcon}\eeq
where $F_Q=\F\cdot\Q/Q$ is the projection of the force onto the
direction of $\Q$.  This has the following interpretation.  If $1/Q$
is regarded as
the length scale of the scattering event, it is required that the
work done by the force $F_Q$ in moving the nucleus this distance be
negligible compared with the root mean square kinetic energy of
the nucleus due to motion along $\Q$; in other words, forces on the
nucleus do not change its energy appreciably during
the scattering process.  This feature of impulsive scattering is
reflected in the energy-conserving delta function in \eq{impsqw},
which would otherwise involve a contribution from the nucleus'
change in potential energy.

An alternative discussion of the short time expansion can be made
within the framework of the Gaussian approximation for incoherent
scattering \cite{egel62}, in which the correlation function
is written $\yqt=\exp[-Q^2w(t)]$, with the width function $w(t)$
proportional to the mean square displacement of the atom at time
$t$.  The impulse approximation is obtained in an asymptotic analysis
by expansion of $w(t)$ about a saddle point in the complex plane.
The correction terms may be estimated in terms of the moments of
$\sqw$, and are small if the ``skewness'' of the recoil peak, related
to the third moment, is small.  The condition for validity of the
\ia\ obtained in this way is equivalent to \eq{selfcon}.

\subsection{Bound nuclei}\label{bound}
The preceding derivation of the conditions for validity of the \ia\
is not rigorous, in that the use of the \ia\ to predict its own
range of validity is a circular argument.  An example in which the
reasoning fails is that of scattering from a nucleus bound in a
harmonic potential.  This example will now be treated in some
detail.  It will lead us to extend somewhat the concept of scattering
time, and will serve as an introduction to a general discussion of
scattering from nuclei bound in solids.

For the harmonic oscillator, $\sqw$ can be calculated without
approximation \cite{love84}.  It will be sufficient for our
purposes to examine the case of zero temperature, for which the
exact result is
\beq
\sqw=\hbar^{-1}e^{-\w_R/\w_0}\sum_{n=0}^\infty[(\w_R/\w_0)^n/n!]
\,\delta(\w-n\w_0).
\eeq
The prediction of the \ia\ is
\beq
S_i^{\hbox{\scriptsize IA}}=(2\pi\hbar^2\w_0\w_R)^{-1/2}
\exp\left[-{(\w-\w_R)^2\over2\w_o\w_R}\right],
\eeq
and the self-consistent estimate of its range of validity, from
\eq{selfcon}, is $\w_R\gg\w_0$.  The two functions are compared in
Fig.~2 for $\w_R/\w_0=15$.  The exact $\sqw$ is an array of infinitely
sharp lines, whereas the \ia\ result is a smooth Gaussian, drawn
as a dashed curve.  The \ia\ fails spectacularly to reproduce the fine
structure of the spectrum.

\def\capttwo{\small
Solid vertical lines: the response function for scattering
from a nucleus bound in an ideal harmonic well of frequency $\w_0$
(each delta function contribution is plotted as a vertical line of
height equal to the corresponding weight).  The momentum transfer
$\Q$ is such that the recoil frequency $\w_R$ is $15\w_0$.  Dashed
curve: the same quantity as predicted by the \ia.}
\begin{figure}
\centerline{\psfig{file=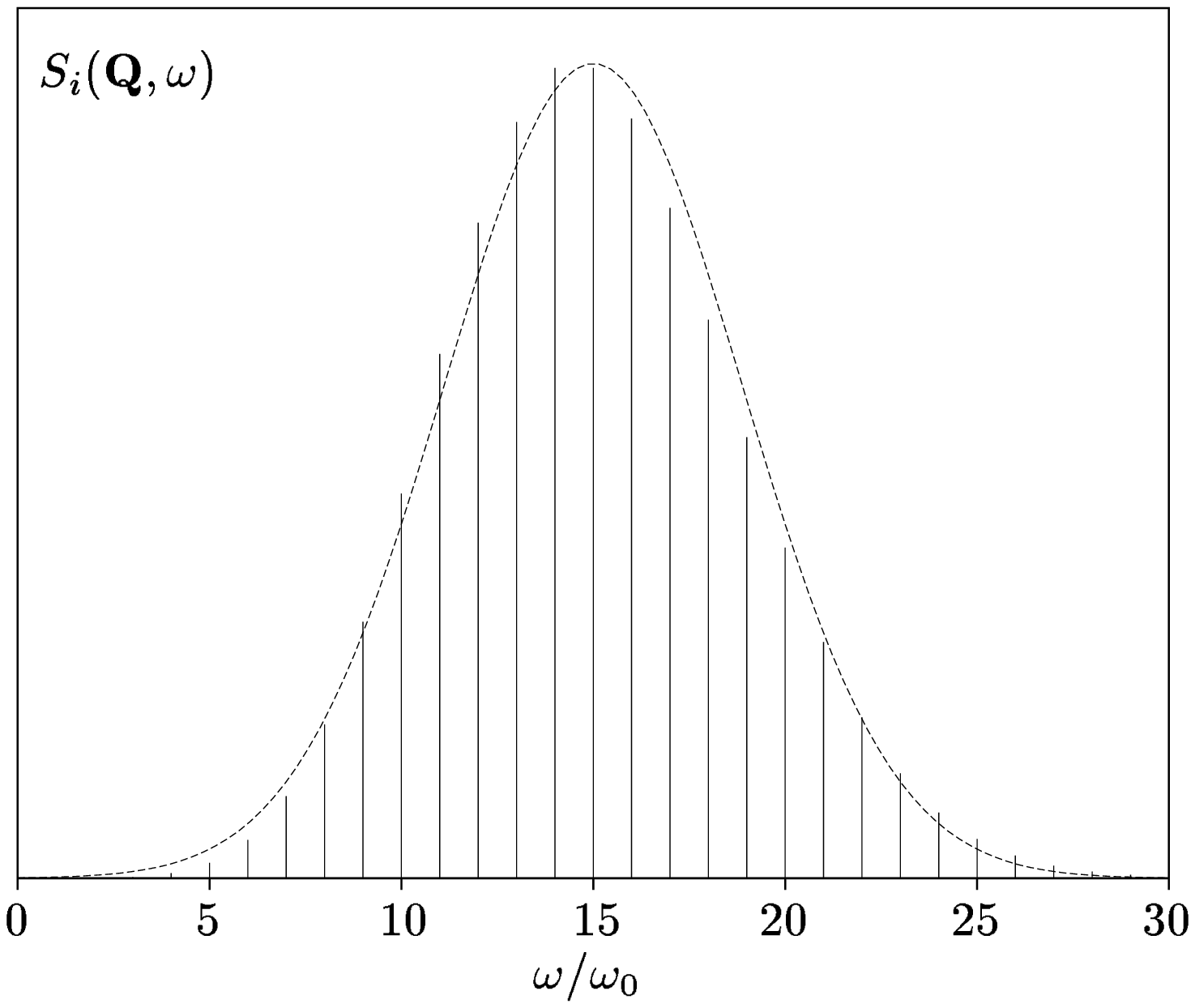,width=4truein}}
\caption{\capttwo}
\end{figure}

As is evident from the general expression, \eq{boundia}, for the
response function in terms of single particle states, the source of
the problem is energy quantization.  Each sharp line represents
the absorption of a number of quanta by the oscillator.  Since $\sqw$
contains infinitely narrow structure, the scattering time, as defined
above, is infinite, and the short time expansion fails.

Not all is lost, however.  Evidently, the impulse approximation does
give an accurate account of the {\it envelope} of the palisade of delta
functions.  The agreement improves on increasing $Q$.  In other words,
the \ia\ describes $\sqw$ on a frequency scale large compared with $\w_0$.
In practice, such a coarse description is likely to be adequate.  For
nuclei bound in molecules or solids, even if broadening mechanisms
intrinsic to the system (discussed below) are insufficient to smear
out structure on a scale of $\w_0$, achieving the instrumental resolution
needed to discern the separate lines would most likely be a difficult task,
with an incident energy large enough to accomplish the experiment.

The ability of the \ia\ to describe the envelope of the discrete
spectrum is not restricted to the ideal oscillator.  The generalization
to arbitrary potential wells is straightforward \cite{gunn86}.  At zero
temperature, \eq{boundia} is
\beq
\sqw=\sum_n|\av{0|e^{-i\Q\cdot\R}|n}|^2\delta(\hbar\w-(E_n-E_0)).
\label{tzero}
\eeq
The scattering intensity is concentrated at values of $\w=(E_n-E_0)/\hbar$
for which the matrix element is large.  Now, the essence of the \ia\ is
that the momentum transfer, and hence the energy transfer, is so large
that the nucleus' final state is approximately that of a free particle,
a plane wave.  Substituting a final state $e^{i\q'\cdot\R}$ with
$E_n=\hbar^2q'^2/2M$ into \eq{tzero} yields \cite{maye90}
\beq
\sqw=\int n(\q)\delta[\hbar\w-\hbar^2(\Q+\q)^2/2M+E_0]\,d\q,
\label{pwia}
\eeq
where $n(\q)$ is the momentum distribution defined previously in
\eq{momft}.  For each $\q$ there is a contribution to $\sqw$ at a
position relative to the recoil frequency
\beq \w-\w_R=\hbar\Q\cdot\q/M+\hbar q^2/2M-E_0/\hbar. \label{econs}\eeq
For sufficiently large $Q$, the first term dominates the others on
the right hand side, and hence \eq{pwia} is essentially identical to
the \ia\ derived using the short time expansion.

In this re-derivation of the \ia, three approximations
were made.  The first was the replacement of the final state by
a plane wave in the evaluation of the matrix element in \eq{tzero}.
Clearly it is sufficient that this approximation be accurate in the
region of space where the ground state wave function is appreciable,
near the centre of the potential well.  That this is indeed the case
follows from the ideas of the WKB approximation \cite{schi68}, in which
the wave function of a bound state is written as a standing wave with
a local wave vector $q'(x)=[2M(E_n-V(x))/\hbar^2]^{1/2}$.  The impulse
limit requires
\beq V(x)\ll E_n\label{crit1} \eeq
within the spatial extent of the ground state, since then $q'(x)$ is
approximately independent of position.  In addition, to ensure that the
WKB wave function is a reasonable approximation of the excited state
we require the fractional change in $q'(x)$ over one wavelength
$2\pi/q'$ to be small \cite{schi68}.  In the present context this
requires $dV/dx\ll q'E_n$, which is a weaker constraint than \eq{crit1}.

The second approximation made in re-deriving the \ia\ was ignoring
the discreteness of the spectrum of final states.  That is, the sum in
\eq{tzero} was replaced by an integral over all momenta.  It is here
that the delta function structure of $\sqw$ is lost, and the procedure
is therefore justified if our aim is to calculate the envelope of the
spectrum.

The final approximation was the neglect of the second and third terms
of \eq{econs}.  Since $E_0-\hbar^2q^2/2M$ is, on average, the potential
energy in the ground state, we obtain the condition
\beq
\av{V}\ll \hbar Q\av{P_Q^2}^{1/2}/M. \label{crit2} \eeq
This is evidently a stronger condition than \eq{crit1} since the right
hand side increases only linearly with $Q$.  Hence, if \eq{crit2} is
satisfied, all the approximations leading to the \ia\ are justified.

The condition may be expressed in terms of the forces on the atom by
noting that $\F=-\nabla V$.  If $\bar F$ represents the root mean
square force in the ground state, which has a spatial extent of
order $\hbar/\av{P^2}^{1/2}$, then the ground state potential energy
is approximately $\hbar\bar F/\av{P^2}^{1/2}$.  Inserting this
estimate in \eq{crit2} and ignoring any dependence on the direction of
$\Q$ yields $\bar F\ll Q\av{P^2}/M$, in agreement with the
self-consistent estimate, \eq{selfcon}.

We conclude that the short time expansion, and the associated
self-consistent assessment of its validity, are adequate for scattering
from a bound nucleus, provided we are satisfied with a description of
the envelope of $\sqw$.  In this context, it is appropriate to define
the scattering time $\t_s$ as the inverse width of the narrowest envelope,
which is the time scale of processes during scattering which determine the
broad structure of the response function.

The present discussion has considered atoms bound in potentials with
infinitely high walls.  In practice, of course, a sufficiently large
impulse will eject the atom from the molecule or lattice which binds it.
It has been suggested in the literature that the \ia\ is valid only when
the recoil energy is large compared with the binding energy $E_B$.  This
is indeed the condition necessary for the scattering response to be a
smooth function as the \ia\ predicts, rather than an array of narrow lines.
However, the conclusion of this section is that there is an intermediate
energy range, $\w_0\ll\w_R <E_B/\hbar$, in which the \ia\ provides an
accurate description of the envelope of the response function.

\subsection{Solids}\label{solids}
We turn now to the subject of scattering from nuclei bound in a lattice,
in which motions of different nuclei are coupled.  In other words, we
wish to generalize the preceding discussion, which treated lattice
vibrations as a collection of independent oscillators (i.e.~in an
Einstein model), to a situation where the vibrational modes are
collective in nature.  We aim to provide qualitative estimates for the
range of $Q$ for which the \ia\ is an accurate description of the
response function.  The task of applying the \ia\ in detailed
calculations for particular model systems will not be attempted here;
see, for example Refs.~\cite{love84} and \cite{gunn84}.

The first point to be made is that the picture of $\sqw$ as an array
of delta function lines is no longer appropriate when atomic vibrations
are coupled.  The reason is simply that final states excited by the
scattering are drawn from a continuous distribution, namely the phonon
\dos, rather than from a set of quantized levels as for the oscillator.
To be specific, let us consider a perfect harmonic lattice, at zero
temperature.  Referring to \eq{tzero}, the final states are now taken
to be excited states of the lattice as a whole, rather than
single-particle states.  It is convenient to group together terms with
a given number of phonons excited.  The leading term, then, has $\w=0$
and yields a delta function elastic line just as for the harmonic
oscillator.  The one-phonon contribution, however, is not a sharp line
but a continuous distribution proportional to the phonon density of
states $Z(\w)$.  Explicitly \cite{love84,gunn84},
\beq
\sqw = e^{-2W(Q)}\left\{\delta(\w)+(\w_R/\w)Z(\w)+\ldots\right\},
\eeq
where $e^{-2W(Q)}$ is the Debye--Waller factor.  Continuing the series,
the two-phonon term is the product of a matrix element $e^{-2W}(\w_R/
\w)^2$ and the two-phonon density of states,
\beq \int_0^\infty Z(\w')Z(\w-\w')\,d\w', \eeq
which is the convolution of $Z$ with itself.  Higher terms in the phonon
expansion are proportional to repeated self-convolutions of $Z$, and
therefore become progressively broader and smoother.  If $Q$ is large,
by the time energy transfers of the order of the recoil energy are
reached, the number of phonons excited is so large that the central limit
theorem applies and the $n$-phonon \dos\ is essentially Gaussian.  The
different contributions merge to form a smooth recoil peak.  These features
are illustrated in calculations by Mayers, Andreani and Baciocco
\cite{maye89} using a Debye model, and by Evans {\it et al.}~\cite{evan96}
and Fielding {\it et al.}~\cite{fiel96} using realistic densities of
states for ZrH${}_2$ and ZrD${}_2$.

The smoothing effect of the continuous distribution of phonon states
is a ``de-phasing'' effect, i.e.~$\yqt$ decays because the spread in
frequencies causes a loss of coherence over time.  This is to be
contrasted with decay of correlations due to damping mechanisms.  The
latter are a consequence of interactions of phonons with impurities,
with electrons, and with each other (due to anharmonicity), resulting
in a finite phonon lifetime.  Such effects may be included in \eq{tzero}
by replacing the delta function by a peak of width equal to the inverse
phonon lifetime.

The effects of finite temperature are to broaden the scattering response
still further.  They enter through both de-phasing and damping
mechanisms: the former because of the $k_BT$ spread in initial energies,
the latter because the phonon lifetime is a decreasing function of
temperature.

Let us summarize the various energy scales that have arisen in our
discussion so far.  The characteristic frequency of the phonons is
the Debye frequency $\w_D$, corresponding to an energy scale typically
of the order of a few hundred kelvin.  For low temperatures,
$k_BT\ll\hbar\w_D$, the mean kinetic energy per atom, $E_K$, is of order
$3\hbar\w_D/4$, while for high temperatures it approaches the classical
value $3k_BT/2$.  The lifetime $\t$ of phonons is related to the
thermal conductivity of the solid by $K=Cv^2/3\t$, where $C$ is the
heat capacity and $v$ is the sound velocity \cite{kitt86}.  Inserting
representative values leads to an estimate that the lifetime
broadening $\hbar/\t$ is typically 1 K or less, and hence negligible
in comparison to other energy scales in the system.

The energy scales related to the scattering process are the recoil
energy $\hbar\w_R$, typically in excess of 1 eV, and the width
$\dw$ of the recoil peak.  These are the scales which determine the
applicability of the \ia.  In view of the smoothing effect of the
continuous phonon \dos, we judge that the self-consistent criterion
for the validity of the \ia\ is appropriate.  In order to apply it,
we require an estimate of the forces on an atom.  For a harmonic
oscillator, the mean square of one Cartesian component of the force is
$M^2\w_0^4\av{x^2}=2M\w_0^2E_K/3$, and thus for a harmonic or nearly
harmonic solid $F_Q$ is of the order of $(M\w_D^2E_K)^{1/2}$.  The
criterion for validity of the \ia\ becomes
\beq
Q \gg (M\w_D^2/E_K)^{1/2} \sim (M\w_D/\hbar)^{1/2}, \label{solcond}
\eeq
where the second estimate is valid for temperatures low compared with
the Debye temperature.  The quantity on the right is the inverse of the
root mean square displacement of a harmonic oscillator of frequency
$\w_D/2$.  Thus, the criterion for the \ia\ is more easily satisfied
if the atoms are weakly bound, as one would expect.  As remarked
previously \cite{reit85a,maye89}, Debye frequencies do not depend strongly
on atomic mass, and hence the $Q$ value required to reach the impulse
limit should increase with mass as $M^{1/2}$.

Neutron Compton scattering investigations of monatomic solids reported
recently include experiments on argon, krypton and xenon \cite{peek91,%
frad94a}, lithium \cite{evan94}, sodium \cite{fult94} and ${}^4$He
\cite{bafi95}.  To take a representative example, lithium \cite{evan94}
has a mass of 7 a.m.u. and a Debye temperature of approximately 400 K,
yielding the criterion $Q\gg8$ \AA${}^{-1}$ for the impulse limit.
For $Q\sim100$ \AA${}^{-1}$, the highest used in the experiment cited,
the recoil frequency $\w_R\sim3$ eV, and the recoil peak width $\dw$ is
about 0.3 eV.

These considerations are readily generalized to more complicated systems.
An example is the hydrogen molecule, which has been the subject of recent
experiments \cite{maye93,andr95,bafi96a,bafi96b}.  In this system the
binding of protons within an H${}_2$ molecule is much stronger than the
forces between molecules.  The mismatch of energy scales is reflected in
the spectrum of vibrational states, which is envisaged as consisting of
two bands: a broad acoustic band, corresponding to molecular motions
characterized by the Debye frequency $\w_D$, and a narrow high frequency
band centred at the intramolecular vibrational frequency $\w_V$.  The
width of the narrow band is of the order of $\w_D^2/\w_V$.  In H${}_2$,
$\w_V$ is nearly two orders of magnitude greater than $\w_D$, and the
molecular and intramolecular motions are effectively decoupled.  The
characteristics of the scattering are thus dominated by the intramolecular
vibrations, and in fact Mayers \cite{maye93} found the scattering from
liquid H${}_2$ to be indistinguishable from that from the polycrystalline
solid.  In the low temperature limit ($k_BT\ll\hbar\w_V$) appropriate
here, the criterion for validity of the \ia\ is $Q\gg(M\w_V/\hbar)^{1/2}
\sim10$ \AA${}^{-1}$, well within the experimental range.

Similar arguments apply for scattering from light atoms in a heavy
lattice \cite{warn83}, such as the proton in KHCO${}_3$ \cite{post91}.
Here the proton vibrational modes are found in a high frequency band,
narrower than the acoustic band by approximately $(M_1/M_2)^{1/2}$,
and are effectively decoupled from other motions because of the
large mass ratio.

\subsection{Liquids}
The case of scattering from liquids is the most well developed and, at
the same time, the most controversial, application of the
\ia.  Here we restrict ourselves to a few general remarks; details
may be found elsewhere \cite{hohe66,sosn91,gers73,wein82,sear84,kirk84,%
plat84,silv87,rina87,silv88a,silv88b,silv89b,snow92}.

For a monatomic classical liquid, such as a condensed heavy noble gas
\cite{peek91}, the application of the impulse approximation appears
straightforward.  For example, taking order of magnitude values for the
force between atoms in the liquid near its triple point, $\av{F^2}^{1/2}
\sim10^3$ K\AA${}^{-1}$ \cite{peek91}, and the kinetic energy
$E_K=3k_BT/2$ in the range 200 to 400 K, we obtain an estimate
$Q\gg5$ \AA${}^{-1}$ for the impulse limit to be reached.  Indeed, in
the experiments cited, deviations from the \ia\ were observed to be
small at momentum transfers between 17 and 29 \AA${}^{-1}$.

The situation appears to be not very different for most monatomic
quantum fluids, except, of course, that the evaluation of the kinetic
energy must take account of the quantum zero point motion of the atoms.
Examples of recent experiments on normal liquids in which quantum effects
are important include studies of ${}^4$He \cite{azua95a,ande94,andr94,%
bafi95}, ${}^3$He \cite{azua95b} and neon \cite{frad94b,azua95a}.

One way to estimate orders of magnitude is to use a ``cell model'' of
the liquid, in which, on short time scales, an atom is assumed to move
in a roughly spherical cage created by neighbouring atoms, with
intermolecular potentials of, say, Lennard--Jones form.  For example,
Andreani \etal~\cite{andr94} have argued that such a model, with
the total interatomic potential represented by an effective harmonic
vibrational frequency $\w_0=14$ K, accounts reasonably well for the
observed temperature dependence of the mean kinetic energy of normal
liquid ${}^4$He.

Assuming that the same model suffices to estimate the forces on an
atom, we find that $Q\gg(M\w_0/\hbar)^{1/2}\sim1$ \AA${}^{-1}$ is the
condition for the impulse approximation to be accurate.  In fact, this
type of estimate is highly misleading.  The impression that the impulse
limit is attained for rather low $Q$ results from the low mass and the
fact that the attractive part of the Lennard--Jones potential is very
weakly binding.  However, the experience gained from a variety of
theoretical and experimental investigations over the past few decades
(see \cite{sosn91} and references therein) has shown that the repulsive
``hard-core''--like part of the interatomic potential is crucial in
determining the validity of the \ia.  Indeed, as mentioned previously,
for a true hard-core fluid the impulse limit is not reached no matter
how large the momentum transfer \cite{reit85b}.

For the most extreme example of a quantum liquid, ${}^4$He in its
normal and superfluid phases, the work of Silver \cite{silv88a,silv88b,%
silv89b} is a definitive theoretical study of corrections to the \ia.
Its conclusion is that, although the correction terms in a formal
expansion of $\sqw$ as $\Q\to\infty$ are proportional to powers
of $1/Q$ \cite{sear84}, the nearly hard-core nature of the interactions
results in a broad range of crossover to the asymptotic limit, in
which the corrections scale as $\log Q$.  As a result, the corrections
are not negligible, even for the highest $Q$ values that might
conceivably be attained in experiments.  Silver's work also provides
a systematic method of calculating the corrections, which has
been applied successfully in measurements of the condensate fraction
in superfluid helium \cite{snow92}.

For non-monatomic liquids, there are complicating (and interesting)
features arising from the internal structure of the molecule.  Recent
work includes experiments on molecular hydrogen and deuterium
\cite{maye93,andr95,bafi96a,bafi96b}.

\subsection{Corrections to the \ia}\label{corr}
Deviations of the response function $\sqw$ from the prediction of
the \ia\ are often collectively termed ``final state effects'', since
they result from deviations of the final state of the scattering
process from plane wave form.  In fact, since $\sqw$ reflects
properties of initial as well as final states, and deviations arise
from both sources \cite{maye90,maye89}, we shall refer simply to
``corrections to the \ia''.

In general, the corrections take the form of a broadening of
the response function; this is in part an effect of a finite
lifetime of the final state, due to collisions of the struck atom
with its neighbours \cite{hohe66}.  An example is the broadening
effect of the phonon lifetime discussed in Sec.~\ref{solids}.  The
corrections are frequently embodied in a ``final state resolution
function'', which is convolved with the result of the \ia\ to
obtain the predicted Compton profile.

When deviations from the \ia\ are appreciable, accurate calculations
of the final state resolution function require a detailed theory of
interactions in the system under study.  An example is the hard-core
perturbation theory \cite{silv88a,silv88b,silv89b} for normal and
superfluid ${}^4$He, which is a quantum many-body theory of the fluid
incorporating scattering data for the helium interatomic potential.
Here, we shall not enter into details of such calculations, but make a
few general comments, principally aimed at the case of scattering from
solids for which the corrections are small.

Let us recall the approach taken in Sec.~\ref{bound} and
Ref.~\cite{gunn86}, where the \ia\ was derived under three
assumptions: that the final state is nearly a plane wave, that the
discreteness of the final state energies is unimportant, and that
the last two terms in the energy conservation condition, \eq{econs},
may be neglected.  The last assumption, that
\beq E_0 \approx \hbar^2q^2/2M, \label{bung}\eeq
is exactly true only for a free particle system.  Mayers \cite{maye90}
has termed the deviations from \eq{bung} ``initial state effects''.
Here we focus on the fact that, as mentioned previously, \eq{bung}
is the strongest of the three approximations made, and therefore if
we relax it, the result is likely to be a better approximation than
the \ia.  For example, if the final state energy $E_n$ is set equal
to its WKB estimate $\hbar^2q^2/2M+V(x)$, and if $E_0-V(x)$ is replaced
with its average value, namely the mean kinetic energy $E_K$, one
obtains
\beq
\sqw=\int n(\q)\delta[\hbar\w-\hbar^2(\Q+\q)^2/2M+E_K]\,d\q,
\eeq
a result first suggested by Stringari \cite{stri87}.  It is a better
approximation than the \ia\ at low temperatures, but becomes less
useful at higher temperature, since the replacement of a distribution
of energies by an average ceases to be valid \cite{maye89}.

The Stringari formula shows that one aspect of the corrections to
the \ia\ is a shift of the maximum of the recoil peak to lower
frequency.  Such a shift is indeed visible in Fig.~2, for example.
In addition, the corrections induce an asymmetry in the peak.  This
suggests that a straightforward symmetrization of the data can be
used to remove partially deviations from the \ia.  In the Sears \cite{sear84}
method, symmetrization is the first step in a systematic self-consistent
correction procedure.  The method is based on a formal asymptotic
expansion of $\sqw$ in powers of $1/Q$, in terms of successive moments
of the final state resolution function.  It is found that symmetrization
of the measured recoil peak eliminates corrections of order $Q^{-1}$,
leaving residual terms of order $Q^{-2}$.  The antisymmetric part of $\sqw$
is proportional to $Q^{-1}$ and amounts to a measurement of deviations
from the \ia; this information can then be used to correct for the residual
deviations in the symmetrized data, and the result is a response function
corrected up to order $Q^{-2}$.  Comparisons of correction methods for
experimental data on bound light atoms have been made by Evans
{\it et al.}~\cite{evan96} and by Fielding {\it et al.}~\cite{fiel96}.

An analysis method recently developed by Glyde \cite{glyd94b}, also
based on moment expansions, aims to extract both the limiting
impulse approximation response function, and measurements of the
finite $Q$ corrections, from experimental \ncs\ data.  In this
approach, rather than concentrating on extreme momentum transfers in
order to minimize deviations from the \ia, one collects data over
a broad range of $Q$.  It is the measurable difference in the $Q$
dependence of various contributions to the expansion moments that
allows the $Q\to\infty$ (\ia) component of the data to be isolated,
and the corrections to be estimated.  This technique has been applied
in a recent study of normal liquid ${}^4$He and liquid neon
\cite{azua95a}.

\section{The Compton profile}
In this section we turn to the problem of extracting information about
the momentum distribution from measured data.  We shall take the
approach \cite{reit85a,reit89} of assuming that the \ia\ is accurate.
As we have seen, this assumption is justified in the case of scattering
from solids, where the impulse limit is well within the experimental
range of momentum transfers and where leading order ``final state''
corrections can be handled, for example, by symmetrization of the data.

The measured response function $\sqw$ is given in the \ia\ by \eq{impsqw}.
Let us define $J=(\hbar^2Q/M)S_i$ and rearrange the delta function to give
\beq
J = \int n(\q)\delta(\q\cdot\Qh-y)\,d\q, \label{radon}
\eeq
where $\Qh$ is the unit vector along $\Q$, and
\beq y=(M/\hbar Q)(\w-\w_R). \label{ydef} \eeq
This form of $\sqw$ has an important consequence.  Consider an isotropic
system, where the scattering is independent of the direction of $\Q$.
Then $J$ depends only on the variable $y$, rather than on $\Q$ and $\w$
independently.  In other words, the data ``collapse'' onto a function
$J(y)$ of a single variable.  This phenomenon is known as $y$-scaling,
and $J(y)$ is termed the \cp.  It is seen from the delta function in
\eq{radon} that $y$ is the projection of the atom's momentum onto the
scattering vector.

The function $J(y)$ is a convenient and standard form for experimental
results to be presented.  The recoil peak is shifted to be centred at
$y=0$, and if the profile is normalized to unity the mean kinetic
energy per atom is directly related to the second moment of the \cp\ by
\beq E_K=(3\hbar/2M)\int_{-\infty}^{\infty}y^2J(y)\,dy. \eeq
This result follows from the general moment relations for the incoherent
response function \cite{love84}.

For an anisotropic system such as a single crystal, the response
function depends on the direction of $\Q$ as well as on $y$, and one
then speaks of the directional \cp, $\jqy$.  It has a second moment
related to the kinetic energy associated with motion along $\Qh$.

\subsection{Reconstructing momentum distributions}\label{recon}
The \cp\ and the momentum distribution are related by \eq{radon}.
Consider $\jqy$ as a function of $y$ for fixed $\Qh$, and define
a coordinate system $xyz$ such that $\Qh$ points along the $z$-axis.
Then
\beq \jqy=\int n(q_x,q_y,y)\,dq_xdq_y. \eeq
The \cp\ is seen here to be the integral of $n(\q)$ over a plane
normal to the vector $\Qh$, i.e.~a projection of the momentum density
onto the direction of the momentum transfer.  This mathematical
relationship between $\jqy$ and $n(\q)$ is known as a Radon
transform \cite{dean83}.  An important property of the transform is
that it is invertible:  given the directional \cp\ for all values of
its arguments, the momentum distribution can be extracted.

The Radon inversion formula can be expressed in many mathematically
equivalent forms, but these are not equivalent in practice, when the
data are finite and affected by noise and instrumental resolution.
The reconstruction problem for the Radon transform, and the associated
mathematical questions of stability, uniqueness, accuracy and resolution,
have been studied thoroughly in connection with computerized tomography
in diagnostic radiology \cite{natt86}.  We note, in passing, that the
Radon transform can be expressed in the framework of wavelet theory
\cite{hols95}, but the consequences, if any, for the reconstruction
problem do not appear to have been explored.  Here we describe a
reconstruction method proposed by Reiter \cite{reit85a,reit89} based on
the work of Davison and Grunbaum \cite{davi81a,davi81b}, which is a
variant of the ``filtered back-projection'' technique in common use in
medical applications of tomography.

The method involves decomposing the angular ($\Qh$) dependence of the
data into spherical harmonics, and the $y$-dependence into products of
Gaussian and Hermite functions.  Specifically,
\beq
\jqy=\pi^{-1/2}e^{-(y/y_0)^2}\sum_{nlm}A_{nlm}H_{2n+l}(y/y_0)
Y_{lm}(\Qh), \label{adef}
\eeq
where $Y_{lm}$ are the usual spherical harmonics, as defined, for
example, in Ref.~\cite{schi68}.  The constant $y_0$ is chosen to match
the width of the \cp.  As may be proved by direct integration
\cite{reit89}, this series is the Radon transform of
\beq
n(\q) = \pi^{3/2}e^{-(q/y_0)^2}\sum_{nlm}B_{nlm}(q/y_0)^lL_n^{l+1/2}
(q^2/y_0^2)Y_{lm}(\hat\q), \label{bdef}
\eeq
where
\beq B_{nlm}=(-1)^n2^{2n+l}n!A_{nlm}. \label{abeq} \eeq
Here $H$ and $L$ are Hermite and Laguerre polynomials, with the standard
normalization \cite{abra65} that the coefficient of $x^m$ in $H_m(x)$
is $2^n$, and the coefficient of $x^m$ in $L_m^\alpha(x)$ is $(-1)^n/n!$.
The method is thus to determine the $\{A\}$ coefficients from the data, to
generate the $\{B\}$ coefficients using \eq{abeq}, and finally to obtain
the reconstructed momentum distribution in the series form in \eq{bdef}.

Two properties of this inversion method make it particularly suitable
for application in \ncs.  The first is that instrumental resolution is
readily accounted for by including the resolution function in the
fitting process used to extract the $A$ coefficients.  In other words,
the functions actually used to fit the data are the convolutions of
those on the right hand side of \eq{adef} with the instrumental
broadening function.  The second desirable property is that the expansion
functions are very well matched to the characteristics of the data.
Indeed, if the underlying binding potential is isotropic and harmonic,
then $\jqy$ is an isotropic Gaussian and only the first term, proportional
to $A_{000}$, is needed.  Even for anharmonic
potentials, the \cp\ is expected to consist of a compact peak, and a
fairly small number of terms in the expansion should suffice.  The $\{A\}$
coefficients, therefore, are an economical description of the data, and
the coefficients with $n>0$ are directly related to the anharmonicity
in the system \cite{reit89}.

The expansion of the \cp\ in \eq{adef} has a definite symmetry.  The
Hermite polynomial and the spherical harmonic each have parity $(-1)^l$,
and hence $J(-\Qh,-y)=\jqy$.  This is a symmetry which is obeyed by
$\jqy$ if the \ia\ is valid.  Measured data, of course, will deviate from
perfect symmetry.  The procedure of fitting the symmetric expansion will
ignore the antisymmetric components in the data, i.e.~it will implicitly
perform a symmetrization.  This has, as a bonus, the effect of removing
leading order corrections to the \ia\ (see Sec.~\ref{corr}).  For
scattering from single crystals, there are likely to be additional
symmetries in the directional \cp\ arising from the point group symmetry
of the lattice site of the struck atom.  In this case, the additional
symmetry may be taken into account by expressing the expansion in terms
of lattice harmonics \cite{reit89}.

As a final point, we discuss scaling of the data.  A uniform scale
factor $y_0$ is already included in \eq{adef}.  Choosing $y_0$ to be
the width of the best Gaussian fit to the \cp\ matches the expansion
to the data, and is likely to minimize the number of coefficients needed.
However, this simple prescription is not adequate for strongly anisotropic
data, where the width of the profile is very different in different
directions.  Here we suggest a general procedure for anisotropic scaling.

There does not appear to be a simple relationship between the inverse
Radon transforms of two functions $\jqy$ differing by a linear
coordinate transformation, i.e.~it does not appear useful to apply a
linear transformation to the data.  Instead, we ask the reverse question
of how $\jqy$ changes when we transform the momentum distribution.  Let
us suppose, then, that the anisotropic distribution $n(\q)$ can be made
reasonably isotropic by a linear scale change
\beq \q\to\q'=(q_x/\a_x,q_y/\a_y,q_z/\a_z), \eeq
for a suitable choice of orthogonal axes $xyz$.  Defining $n'(\q)=n(\q')$
and
\beq \Q'=(\a_xQ_x,\a_yQ_y,\a_zQ_z), \eeq
we find the Radon transform of $n'$ to be
\beq
J'(\Qh,y)\propto\int n(\q')\delta(\q'\cdot\Q'-y)\,d\q'
=(1/Q')J(\Qh',y/Q'), \label{anis}
\eeq
where $Q'$ is the length of $\Q$ and $\Qh'=\Q'/Q'$, and an irrelevant
constant factor has been omitted.  The suggested procedure is then as
follows.  Given a measured \cp, $\jqy$, let $\a_x$, $\a_y$ and $\a_z$
be its characteristic widths along the chosen axes.  Compute the
transformed profile $J'(\Qh,y)$ according to \eq{anis}, which amounts
to a nonlinear coordinate transformation.  After applying the expansion
technique described previously to reconstruct $n'(\q)$ from $J'$, use
\beq n(\q)=n'(\a_xq_x,\a_yq_y,\a_zq_z) \eeq
to find the required momentum distribution corresponding to the original
$\jqy$.  The transformed \cp\ $J'$ is roughly isotropic.  For example, if
$\Qh$ points along the $x$-axis, then $J'(\Qh,y) \propto J(\Qh,y/\a_x)$,
which has unit width in $y$, by construction.  Thus $J'$ is a function for
which the Radon inversion step, using the expansion method, is efficient
and stable.

\subsection{Anharmonic potentials}
The Compton profile for scattering from a harmonically bound atom is
Gaussian, and a measurement of its width yields the mean kinetic energy.
Essentially the same information might be obtainable by, for example,
vibrational (infrared) spectroscopy.  Where the \ncs\ technique for
bound atoms comes into its own, then, is in the potential to measure
the anharmonicity of the binding potential, in cases where it is legitimate
to treat the struck atom in terms of single-particle motion in an
effective potential.  It is the only technique capable of measuring
Born--Oppenheimer potentials directly.

The extraction of the potential from the momentum distribution is simply
a matter of inverting the Schr\"odinger equation:
\beq V(\R)-E=(\hbar^2/2M)\psi^{-1}\nabla^2\psi, \eeq
where $\psi(\R)$ and $\psi(\q)$ are Fourier transforms of each other, and
the latter is obtained from $n(\q)=|\psi(\q)|^2$.  The phase ambiguity in
extracting the momentum space wave function from $n(\q)$ is not a problem
for inversion symmetric potentials, since $\psi(\q)$ can always be chosen
to be real.

It should be noted that the technique, at least in its simplest form, is
restricted to systems in which all atoms of the species under study have
identical chemical environments.  If they do not, the Compton profile
will be a superposition of contributions from different environments.

An extreme case of an anharmonic potential is one with two minima, such as
might be expected for a proton in a hydrogen bond \cite{reit85a,reit89}.
A useful model for illustrating the results expected for a double-well
potential is to assume a ground state consisting of two shifted Gaussians
of equal amplitude,
\beq \psi(x)=e^{-(x-a)^2/2\s^2}+e^{-(x+a)^2/2\s^2}, \eeq
in one dimension \cite{reit85a,reit89,maye95}.  This function is plotted
in Fig.~3 for selected values of the ratio $a/\s$, and the corresponding
potentials $V(x)-E$ appear in Fig.~4.  For the largest value of $a/\s$ the
potential consists of rather isolated harmonic wells, which gradually merge
as $a/\s$ is reduced.  For $a<\s$, the potential is more accurately
described as a single harmonic well with a shallow ``bump'' in the centre,
and the wave function is a single non-Gaussian peak.

\def\captthree{\small
Model wave functions for an atom in a double-well potential.  Each is the
sum of two Gaussian components of width $\s$, centred a distance $2a$
apart.  Solid line: $a/\s=0.25$; dashed line: $a/\s=0.6$; dotted line:
$a/\s=1$.}
\begin{figure}
\centerline{\psfig{file=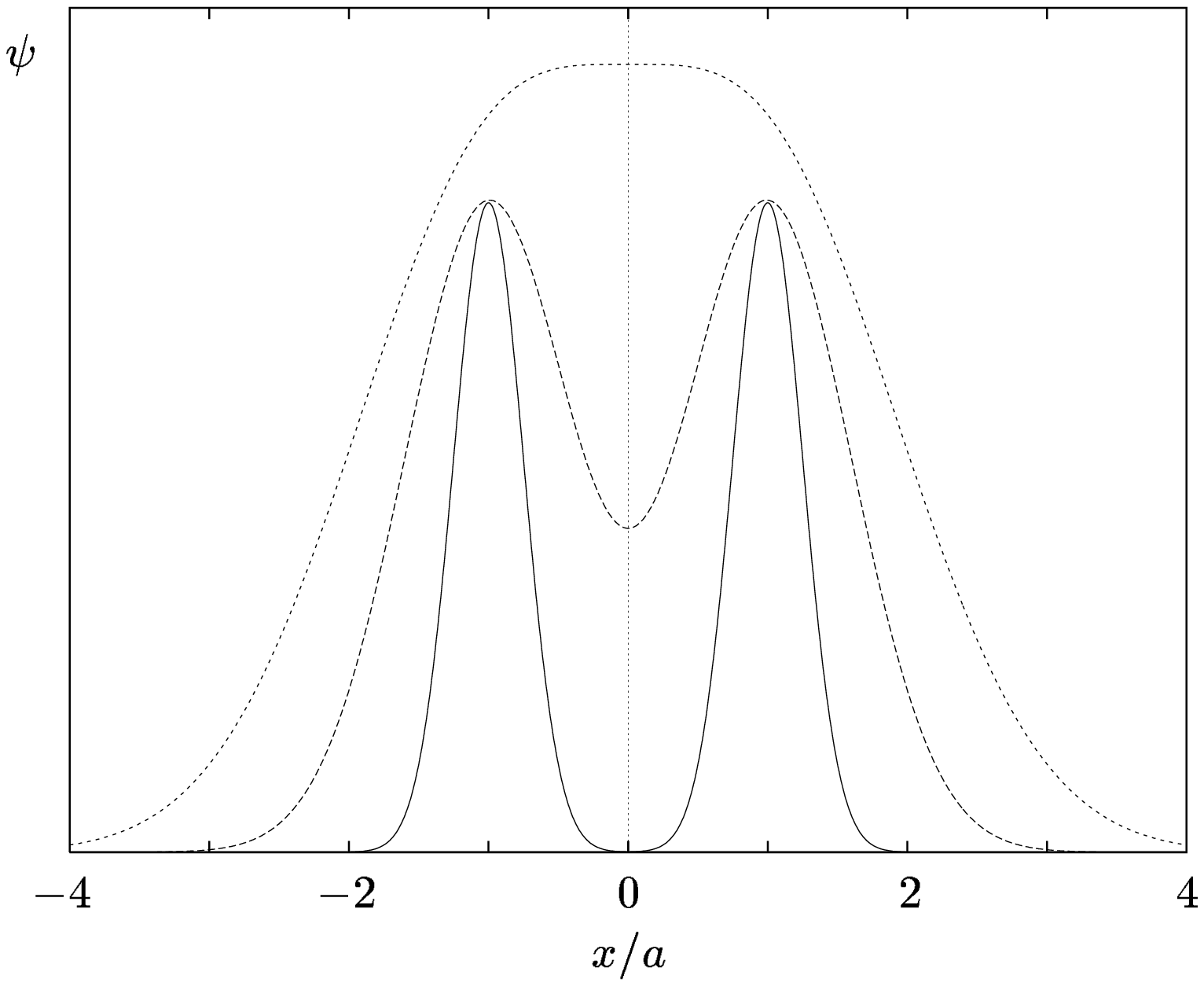,height=3truein}}
\caption{\captthree}
\end{figure}

\def\captfour{\small
Model potentials $V(x)-E$ for each of the model wave functions in Fig.~3.}
\begin{figure}
\centerline{\psfig{file=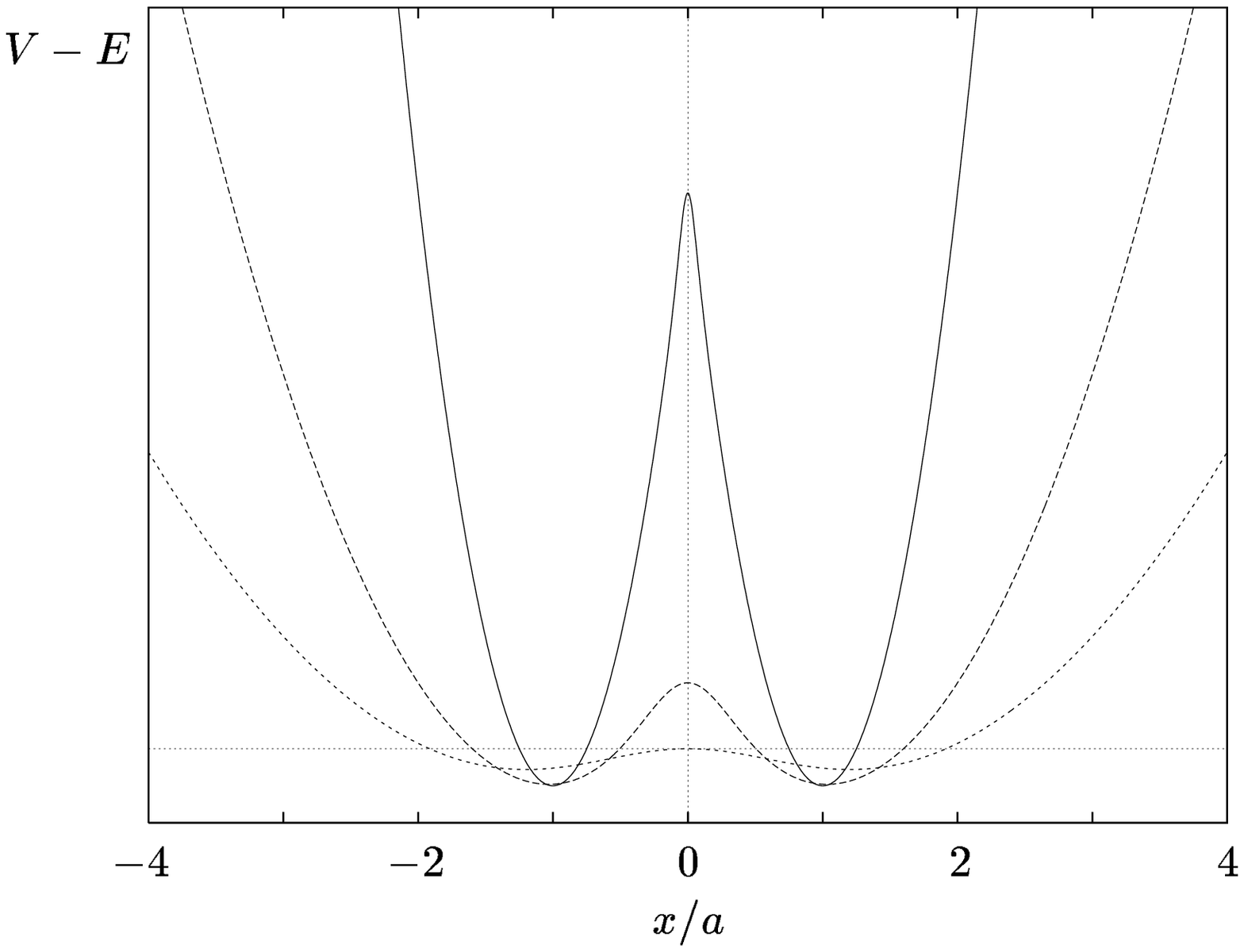,height=3truein}}
\caption{\captfour}
\end{figure}

The \cp, $J(y)=e^{-\s^2y^2}\cos^2ay$, is plotted in Fig.~5 for an
intermediate value $a/\s=0.6$.  It includes an oscillatory factor,
representing interference between wave functions localized in the two
wells, and the overall shape is far from Gaussian.  The number of
oscillations in each half-width of the Gaussian envelope is roughly
$a/\s$, so for weak anisotropy ($a/\s<1$) the deviations from Gaussian
form are less pronounced.  The coefficients in the Hermite expansion,
\eq{adef}, corresponding to this $J(y)$ are $A_n\propto(-a^2/4\s^2)^n/
(2n)!$, which fall off rapidly even for intermediate $a/\s$.  In the
limit of large $a/\s$, the oscillations are so rapid that one observes
only the average, which is Gaussian as one would expect for isolated
harmonic well.

This simple model may be extended to the case where the double well
potential, as in a typical hydrogen bond, is not symmetric.  It is found
that the asymmetry supresses somewhat the oscillatory component in the
wings of the Compton profile, which no longer goes to zero at its
minima.

\def\captfive{\small
\cp\ corresponding to the two-Gaussian model wave function, with
$a/\s=0.6$.}
\begin{figure}
\centerline{\psfig{file=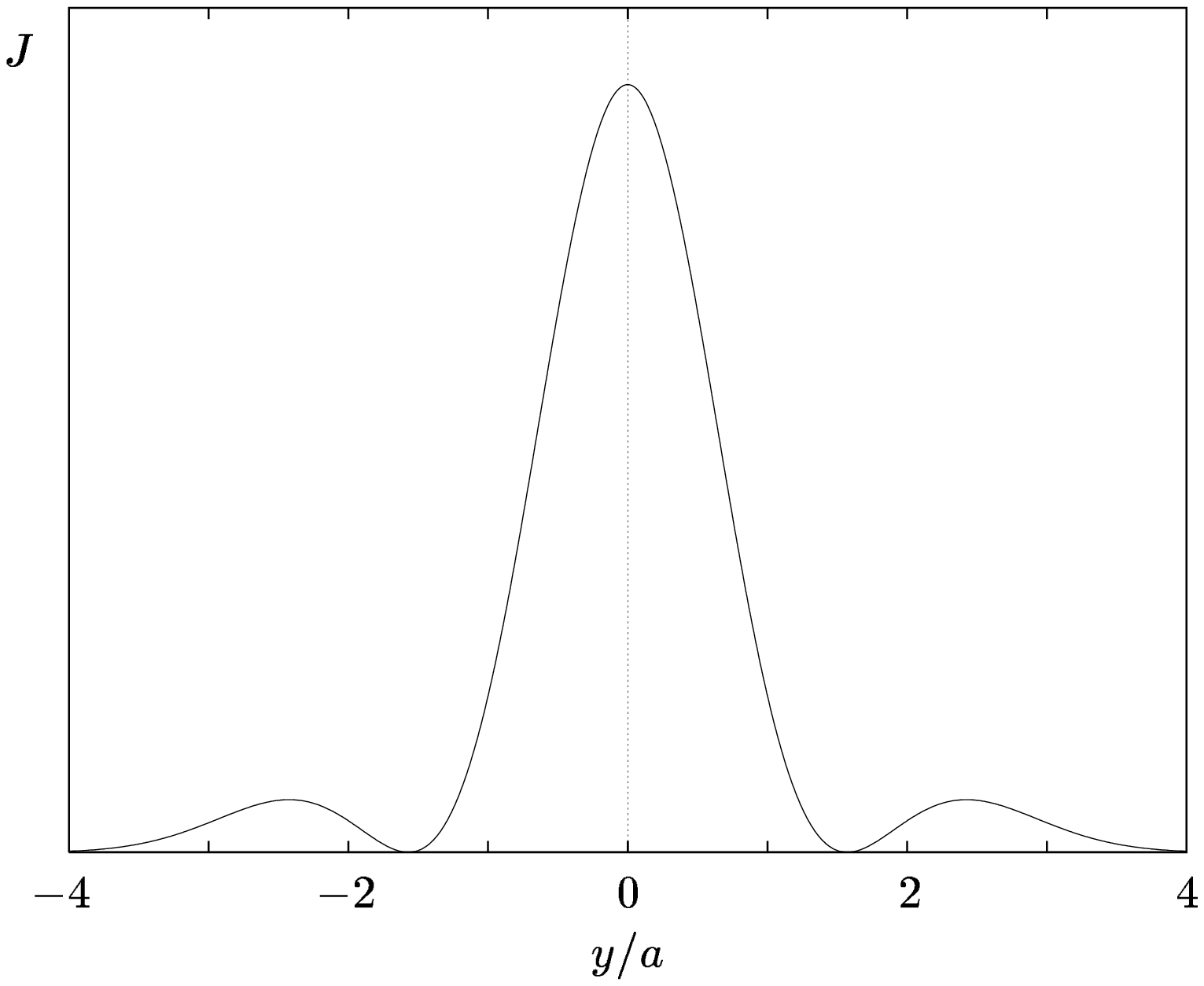,height=3truein}}
\caption{\captfive}
\end{figure}

\section{Discussion}
The \cp\ in Fig.~5, corresponding to a model of an atom in a double-well
potential, is far from Gaussian, which strongly suggests that anharmonicity
of this kind should be experimentally measurable.  Reiter \cite{reit85a,%
reit89} has carried out an analysis of the practical limitations on
measurements of anharmonicity, including the effect of sampling noise (but
not of nonzero instrumental resolution), concluding that extracting
meaningful atomic potential energy functions, using the series expansion
method to analyze data, is feasible.  On the other hand,
Sivia and Silver \cite{sivi89}, in an analysis based on Bayesian probability
theory, have shown that the reconstruction of the momentum distribution
from the \cp\ is an intrinsically ill-posed problem, at least for studies
of the condensate fraction in superfluid helium.  A simple statement of
the essence of their result is that widely different momentum distributions
can yield the same \cp\ within experimental error.  This does not, of
course, imply that momentum distributions cannot be extracted.  There are
grounds to believe that the inverse problem for strongly anisotropic
data for single crystals is better behaved than for liquid ${}^4$He: if
the atomic motion is approximately decoupled into independent motion
along three different axes, the resulting one-dimensional inversions
are more stable than the three-dimensional inversion for isotropic data.
Clearly, further study of the reconstruction problem for \ncs\ from solids
is desirable, to establish, for example, which features of the \cp\ are most
sensitive to anharmonicity, the limits to accuracy of reconstruction, and
optimum algorithms.

Another question that deserves further investigation is the inclusion of
corrections to the \ia\ in the reconstruction procedure.  This is a
well-studied problem for scattering from superfluid helium \cite{sosn91},
while for solids it is argued, as in Sec.~\ref{recon}, that the corrections
are negligible. Nevertheless, we note that in Fig.~5 it would appear to be
the tails of the profile which are most sensitive to the presence of
anharmonicity in the potential, and even if corrections to the \ia\ are
small relative to the recoil peak amplitude, they may be significant in
the tails.

The past achievements of the \ncs\ technique are impressive, ranging from
the mature body of work on superfluid helium, to more recent applications
in studies of quantum and classical liquids, and kinetic energies, anisotropy
and quantum effects in solids.  Here we have summarized the theoretical
background, and the prospects for a new development of the technique, the
reconstruction of momentum distributions by direct inversion from the \cp.
Taking into account theoretical work to date, there are grounds for being
cautiously optimistic.  We await the first direct experimental measurement
of a three-dimensional atomic potential energy function in a solid.

\section*{Acknowledgements}
I would like to thank M.~Celli, A.~L.~Fielding, S.~W.~Lovesey, J.~Mayers,
A.~S.~Rinat, P.~Schofield and D.~S.~Sivia for very helpful discussions.

\end{document}